\documentclass[fleqn,usenatbib]{mnras}

\usepackage[T1]{fontenc}

\DeclareRobustCommand{\VAN}[3]{#2}
\let\VANthebibliography\thebibliography
\def\thebibliography{\DeclareRobustCommand{\VAN}[3]{##3}\VANthebibliography}

\usepackage{graphicx}	
\usepackage{amsmath}	
\usepackage{amssymb}	
\usepackage{ulem}       

\newcommand*{\thead}[1]{\multicolumn{1}{c}{#1}}

\usepackage{newtxtext,newtxmath} 

\title[Conductivity in Abell~2146]{Constraints on thermal conductivity in the merging cluster Abell~2146}

\author[A. Richard-Laferri\`{e}re et al.]{A. Richard-Laferri\`{e}re,$^{1}$\thanks{E-mail: ar999@cam.ac.uk (ARL)}, 
H. R. Russell,$^{2}$
A. C. Fabian,$^{1}$ 
U. Chadayammuri,$^{3}$ 
C. S. Reynolds,$^{1,4,5}$ 
\newauthor
R. E. A. Canning,$^{6}$ 
A. C. Edge,$^{7}$ 
J. Hlavacek-Larrondo,$^{8}$ 
L. J. King,$^{9}$ 
B. R. McNamara,$^{10,11}$ 
\newauthor
P. E. J. Nulsen$^{3,12}$ and
J. S. Sanders$^{13}$\\
$^{1}$Institute of Astronomy, University of Cambridge, Madingley Road, Cambridge CB3 0HA, UK\\
$^{2}$School of Physics \& Astronomy, University of Nottingham, University Park, Nottingham NG7 2RD, UK\\
$^{3}$Centre for Astrophysics | Harvard and Smithsonian, 60 Garden Street, Cambridge, MA 02143, USA\\
$^{4}$Department of Astronomy, University of Maryland, College Park, MD 20742-2421, USA \\
$^{5}$Joint Space-Science Institute (JSI), College Park, MD 20742-2421, USA \\
$^{6}$Institute of Cosmology and Gravitation, University of Portsmouth, Portsmouth PO1 3FX, UK\\
$^{7}$Department of Physics, Durham University, Durham DH1 3LE, UK \\
$^{8}$D\'{e}partement de physique, Universit\'{e} de Montr\'{e}al, Succ. Centre-Ville, Montr\'{e}al, H3C 3J7, Canada\\
$^{9}$Department of Physics, University of Texas at Dallas, 800 W Campbell Rd, Richardson, TX 75080, USA\\
$^{10}$Department of Physics and Astronomy, University of Waterloo, Waterloo, ON N2L 3G1, Canada\\
$^{11}$Perimeter Institute for Theoretical Physics, Waterloo, ON N2L 2Y5, Canada\\
$^{12}$ICRAR, University of Western Australia, 35 Stirling Hwy, Crawley, WA 6009, Australia\\
$^{13}$Max-Planck-Institut f\"{u}r extraterrestrische Physik, Gießenbachstraße 1, D-85748 Garching, Germany
}

\date{Accepted 2023 October 2. Received 2023 September 26; in original form 2022 September 21}

\pubyear{2023}

\begin{document}
\label{firstpage}
\pagerange{\pageref{firstpage}--\pageref{lastpage}}
\maketitle

\begin{abstract}
The cluster of galaxies Abell~2146 is undergoing a major merger and is an ideal cluster to study ICM physics, as it has a simple geometry with the merger axis in the plane of the sky, its distance allows us to resolve features across the relevant scales and its temperature lies within Chandra’s sensitivity. Gas from the cool core of the subcluster has been partially stripped into a tail of gas, which gives a unique opportunity to look at the survival of such gas and determine the rate of conduction in the ICM.  We use deep 2.4~Ms \textit{Chandra} observations of Abell~2146 to produce a high spatial resolution map of the temperature structure along a plume in the ram-pressure stripped tail, described by a partial cone, which is distinguishable from the hot ambient gas.  Previous studies of conduction in the ICM typically rely on estimates of the survival time for key structures, such as cold fronts.  Here we use detailed hydrodynamical simulations of Abell~2146 to determine the flow velocities along the stripped plume and measure the timescale of the temperature increase along its length. We find that conduction must be highly suppressed by multiple orders of magnitude compared to the Spitzer rate, as the energy used is about 1\% of the energy available.  We discuss magnetic draping around the core as a possible mechanism for suppressing conduction.
\end{abstract}




\section{Introduction}

Clusters of galaxies are the most massive gravitationally collapsed objects and their  mergers are the most energetic events in the Universe since the Big Bang.  Mergers dissipate as much as $10^{64}~$ergs of kinetic energy via shocks, turbulence and reacceleration of relativistic particles \citep[e.g.][]{feretti_observational_2002,sarazin_physics_2002}. Galaxy clusters are made up of hundreds to thousands of galaxies, dark matter and hot X-ray emitting gas called the intracluster medium (ICM). During a merger between two clusters, the gas in their ICM collides and is shocked, heated and compressed. Cold fronts, very sharp curved surfaces in three-dimensions characterised by an abrupt inverse jump in density and temperature, can also be created  (for a review on cluster mergers, shock and cold fronts, see \citealt{markevitch_shocks_2007}). Merger cold fronts are discontinuities between cool regions of gas moving at subsonic or transsonic speed through another hotter region \citep[e.g][]{markevitch_chandra_2000}.

Abell 2146 (A2146 hereafter) is the merger of two clusters \citep[][]{russell_chandra_2010,russell_shock_2012,russell_structure_2022}, with the merger axis close to the plane of the sky ($13-19\degr$, \citealt{canning_riding_2012,white_dynamical_2015}). With two clear shock fronts and the post-shock temperatures measurable with \textit{Chandra}, it is the ideal cluster to study ICM physics. Both clusters in this merger were previously cool core clusters \citep{chadayammuri_constraining_2022}, which are clusters that exhibit highly peaked central X-ray brightness distributions and short central radiative cooling times. The remnant of the core of the primary cluster can be observed as a plume in the cluster, while the subcluster's cool core is partially intact. The $\sim2~$keV gas of the subcluster core has been partially ram-pressure stripped from the core and forms a tail of gas that increases in temperature with distance, up to $\sim8~$keV. Looking at the survival of such gas and how it is heated as it gets farther from the core allows for a detailed study of thermal conduction.

Thermal conduction causes heat to flow down a temperature gradient, which for unmagnetized ionized gases follows the Spitzer conductivity equation \citep[][]{spitzer_physics_1956}. It is an obvious way in which the cool gas from the subcluster core can be heated by the hot ambient gas to increase the temperature of the core, and over time destroy the cool core. It has long been discussed theoretically in the case of gas-rich galaxies moving through the ICM \citep[e.g.][]{cowie_thermal_1977}. Conduction is difficult to constrain generally in clusters of galaxies and only a small number of studies on the subject have been possible. \citet{ettori_chandra_2000} carried out the first such study based on the sharp temperature discontinuities observed across the interface of the cold fronts of Abell~2142. They found that conduction needed to be strongly suppressed compared to the Spitzer rate, possibly by magnetic fields and magnetic draping, a phenomenon not well understood. The same conclusion was found in other similar studies (e.g. using the temperature of a filament, see \citealt{fabian_chandra_2001}; using small scale structures in the temperature map, see \citealt{markevitch_chandra_2003}), but all of them used approximations for the survival time of the structures.

In this paper, we use new deep-exposure \textit{Chandra} data of A2146 to obtain high spatial resolution temperature measurements of the  cool core of the infalling subcluster. This cool core is partially offset from the brightest galaxy of the subcluster, traced by the AGN, possibly as a result of the slingshot effect  \citep{hallman_chandra_2004,mathis_formation_2005,ascasibar_origin_2006}.  The cool core is mostly surrounded by a large, sharp temperature gradient. This steep  temperature gradient and the Spitzer conductivity dependence on the 5/2 power of the temperature mean that without any suppression the cool core would be rapidly evaporated.  

We focus on a plume on the eastern edge of the core, which has a clear temperature gradient along its length.  A similar structure is seen in hydrodynamical simulations of A2146 and we use the gas velocities from these simulations to determine the exposure time of the cool gas to the surrounding hot gas. The rate of temperature increase within the plume, if due to thermal conduction, yields robust constraints on the level of suppression. If  mixing is more important, then we place a strong upper limit on the effective conductivity.

In Section \ref{sec:obs}, we present the new 2~Ms of \textit{Chandra} observations of A2146 and the data reduction. In Section \ref{sec:image}, we present the exposure-corrected image of A2146 and some image manipulations to reveal more structures in the gas, while the results on the temperature maps and the study of conduction using the simulation are presented in Section \ref{sec:results}. The implications of the results are discussed in Section \ref{sec:discussion} and the summary is presented in Section \ref{sec:conclusion}. Detailed analysis of the shock fronts is done in \citet{russell_structure_2022}, while this work focuses on the subcluster core. We adopt a $\Lambda$CDM cosmology with H$_0 = 69.6$ km s$^{-1}$ Mpc$^{-1}$, $\Omega_\textnormal{M} = 0.286$, and $\Omega_{\Lambda} = 0.714$ \citep{bennett_1_2014}  throughout this manuscript. The redshift of A2146 is $z = 0.234$ ($1\arcsec = 3.753~$kpc; \citealt{struble_compilation_1999,bohringer_northern_2000}). All errors are $1\sigma$ (68.3\%) unless otherwise noted.

\section{Observations}\label{sec:obs}

New \textit{Chandra} observations of A2146 were obtained between June 2018 and August 2019 with the ACIS-I detector, and have an exposure time of 1.93~Ms over 67 observations (PI Russell). These were combined with the existing ACIS-I observations taken in 2010 \citep[][]{russell_shock_2012}, totalling an exposure time of 2.31~Ms and 75 observations (see \citealt{russell_structure_2022} for the technical details of the observations). Data reduction was done following the standard procedure using \textsc{ciao} v4.13 and \textsc{caldb} v4.9.4 \citep[][]{fruscione_ciao_2006}, with the latest calibration measurements and updates to the ACIS contaminant model, which were crucial for this project (see \citealt{russell_structure_2022} for a study on the calibration of the ACIS gain). It was concluded that the calibration with those versions were sufficient for our analysis. After reprocessing the datasets with \textsc{chandra\_repro}, we applied the improved background screening provided by VFAINT mode. Background light curves were extracted from the diagonally opposite ACIS-I CCD and filtered using the \textsc{lc\_clean} routine to identify and remove periods affected by flares. Only two observations had significant flares (ObsIDs 20921 and 21674), and therefore the resulting exposure time is 2.3~Ms.

We corrected the absolute astrometry of each observation by cross-matching point sources to the ones in ObsID 21733. The event files were also reprojected to match the position of this ObsID as it was reasonably deep and had no calibration issues. For each observation, exposure maps were then created using energy weights determined from an \textsc{apec} model using a redshift of 0.234, a metallicity of $0.319 \pm 0.006~$Z$_{\odot}$ (relative to solar abundances defined by \citealt{anders_abundances_1989} for comparison with past cluster's studies), at the global cluster temperature of $6.67 \pm 0.03~$keV and with the absorption fixed at the Galactic value, $n_{\textrm{H}}= 3.0 \times 10^{20}~$cm$^{-2}$ \citep{kalberla_leidenargentinebonn_2005}. Blank sky backgrounds from ACIS were generated for each observation. Each of them was reprocessed, reprojected to the corresponding sky position using the aspect solution of the observation, and normalized to match the count rate in the $9.5-12~$keV energy band. The accuracy of those blank sky background files was verified by extracting a spectrum from an on-chip region without cluster emission. By comparing both spectra and doing this for multiple observations, it was found that they matched sufficiently to use the ACIS blank sky background for our analysis.

\section{Results}\label{sec:results}

\subsection{Image Analysis}\label{sec:image}

\subsubsection{X-ray image}

\begin{figure}
\centering
\includegraphics[width=0.48\textwidth]{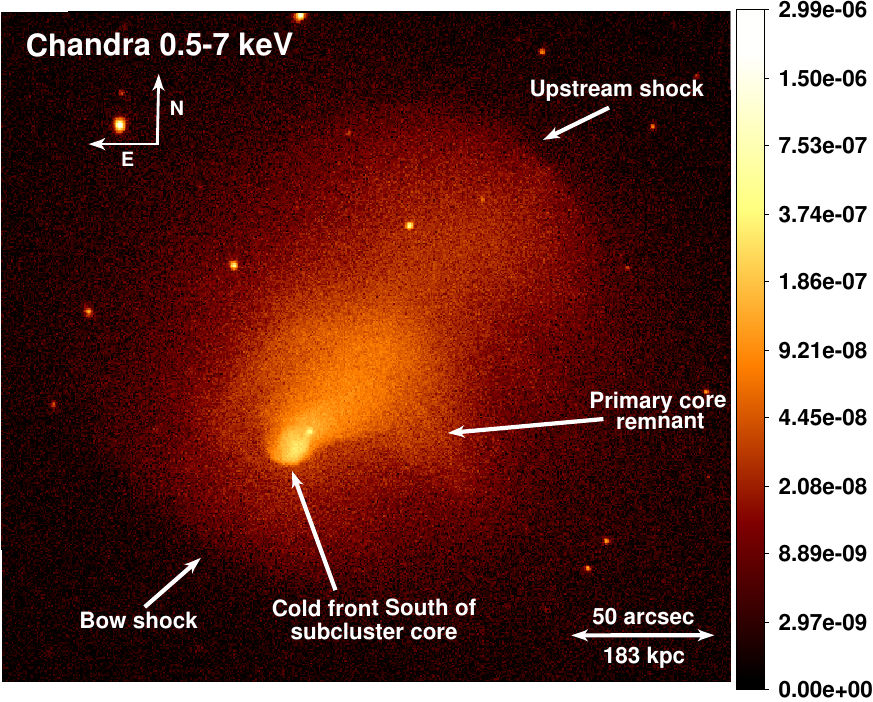}
\caption[]{Exposure-corrected X-ray image of Abell~2146 in the broad energy band ($0.5-7~$keV, surface brightness in counts~cm$^{-2}$~s$^{-1}$~pixel$^{-1}$). We highlighted some of the interesting features. The north direction is up and the east direction is towards the left.}
\label{fig:A2146image}
\end{figure}

The exposure-corrected image in the $0.5-7~$keV energy band, combining all 75 observations, can be seen in Fig.~\ref{fig:A2146image}. It represents a merger between a smaller cluster, identified with the remnant subcluster core, the brightest region in the cluster, and the primary cluster, which was strongly disrupted during the merger \citep{russell_chandra_2010,russell_shock_2012}. The bright region behind the northern shock (upstream shock) consists of shocked gas that fell in from the wake of the smaller subcluster. The plume on the west side of the subcluster core is the remnant of the primary cluster's cool core (see also \citealt{russell_structure_2022}) that was pushed there during the merger. With hydrodynamical simulations, \citet{chadayammuri_constraining_2022} found that it is an off-axis merger, with the two cluster cores passing within $\sim 100~$kpc of one another at closest approach, explaining the lack of symmetry about the merger axis.

The two shock fronts are at opposite sides of the cluster. The bow shock to the south is formed when the subcluster infall velocity exceeds the sound speed in the ICM. On the other hand, the upstream shock is created when gas from the subcluster falling in from the wake of its core runs into gas that has been stripped from it \citep[e.g.][]{roettiger_numerical_1997,gomez_cooling_2002}. The Mach numbers of the bow and upstream shocks are respectively $M=2.44\pm0.17$ and $M=1.58\pm0.05$ \citep{russell_structure_2022}. Combining this with galaxy distribution \citep[][]{canning_riding_2012,white_dynamical_2015} and hydrodynamical simulations, it was found that the subcluster passed through the primary cluster $\sim 0.3-0.5~{\rm Gyr}$ ago and the mass ratio between the two clusters is $3:1$ \citep{chadayammuri_constraining_2022}. We refer the reader to \citet{russell_structure_2022} for a detailed analysis of the shock fronts and the transport processes around them.

The remnant core of the subcluster has survived  passage through the core of the main cluster. However, some gas from the subcluster core has been stripped away by ram-pressure, forming a wide tail of gas extending over 200~kpc to the northwest. We use gas stripping as a general term for ram-pressure pushing on the leading edge of the subcluster core and continuous stripping around the sides, including by Kelvin-Helmholtz instabilities (see e.g. \citealt{roediger_stripped_2015}).  The direction of the stripped tail indicates that the subcluster core is currently travelling to the southeast. This gas flow mimics the one around a solid body and is captured by the simulations \citep[][]{chadayammuri_constraining_2022}. Finally, there is a cold front at the southwest part of the subcluster core.

\subsubsection{Cool core structure and optical image}

\begin{figure*}
\centering
\begin{minipage}[c]{1.0\linewidth}
\centering
\includegraphics[width=1.0\textwidth]{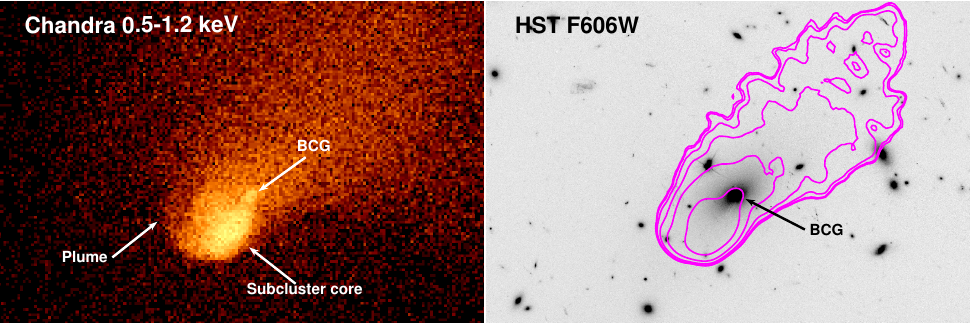}
\end{minipage}
\caption[]{Images on the same scale centred on the subcluster core of A2146. \textbf{Left:} Exposure-corrected X-ray image of Abell~2146 in the soft energy band ($0.5-1.2~$keV, surface brightness in counts~cm$^{-2}$~s$^{-1}$~pixel$^{-1}$) zoomed on the subcluster core. We highlighted some of the interesting features. \textbf{Right:} \textit{Hubble Space Telescope} F606W optical band image showing the subcluster galaxies \citep[][]{king_distribution_2016}. Seven X-ray contours (in magenta) are overlaid on the optical image: from $2.5 \times 10^{-8}$ to $2.5 \times 10^{-7}$~counts~cm$^{-2}$~s$^{-1}$~pixel$^{-1}$. The contours represent the soft band X-ray emission from the left panel. We highlighted the BCG of the subcluster core. For all panels, the north direction is up and the east direction is towards the left.}
\label{fig:unsharpmask_optical}
\end{figure*}

In the left panel of Fig.~\ref{fig:unsharpmask_optical}, there is a zoom on the subcluster core at soft energy bands ($0.5-1.2~$keV), where we can see a break up of the gas from the core on its east edge in contrast to the narrow southwest edge. The southwest side of the core has a sharp edge that extends for $\sim$150~kpc whilst the northeast side has no clear edge and is instead breaking up into a plume or stream of cool gas blobs (see e.g. \citealt{russell_shock_2012}).  This contrast is likely due to the off-axis nature of the merger and corresponding asymmetries in the magnetic field structure and suppression of turbulent instabilities (see \citealt{chadayammuri_turbulent_2022}).  Although we cannot be certain of the 3D-geometry, for simplicity, we refer to this stripped structure on the northeast edge of the core as a plume and assume a conical shape.  In Fig. \ref{fig:unsharpmask_optical}, we highlight this structure along with the subcluster core and its BCG hosting an actively accreting AGN. This plume lies on the outer edge of the tail, therefore projection effects are minimised, and a comparable structure is visible in the hydrodynamical simulations.  This structure is therefore the focus of this paper as a probe of thermal conduction. 

In the right panel of Fig.~\ref{fig:unsharpmask_optical}, we present the optical image of A2146 from the \textit{Hubble Space Telescope} archive, using the F606W optical band, on the same scale as the X-ray image. Using the overlaid soft band X-ray contours, we can clearly see the subcluster's BCG in the optical image, but no galaxies are at the position of the X-ray plume studied. Therefore, the X-ray emission is from the gas of the subcluster core.

\subsection{Spectral Analysis and temperature maps}\label{sec:Tmaps}

Detailed maps of the gas temperature were created to measure the temperature structure of cool plumes ($2-5$~keV) that could be differentiated from the hot ambient background.  We  use the Contour Binning algorithm to extract spectra from small regions \citep[][]{sanders_contour_2006}\footnote{From \url{https://github.com/jeremysanders/contbin}}. This algorithm generates spatial regions of a chosen signal-to-noise by grouping neighbouring pixels of similar surface brightness. The dimensions of the regions can be restricted, and were chosen so that the length was, at most, two and a half times the width. We focused on the central $0.9\arcmin \times 0.9\arcmin$ region and excluded the AGN.

For each observation, a spectrum was extracted from each region using the \textsc{dmextract} tool of \textsc{ciao} and combined with its response files. We used the \textsc{grppha} tool of HEASoft to group each spectrum such that it contained a minimum of one count per spectral channel. To have sufficient counts in spectra extracted from the ACIS background event files, we used a simple blank sky background on-axis region of $2\arcmin \times 2\arcmin$ for each observation, instead of a distinct blank sky background for each region. This implies that there is some correlation between the properties derived for these regions. However, it is important to note that the background subtraction is not particularly significant for the bright core structure analysed here. We refer to \citet{russell_structure_2022} for a detailed analysis of the background for these datasets.

Following this, the spectral analysis of each region was done using \textsc{xspec} version 12.11.0 \citep[][]{arnaud_xspec_1996}. For each region, the spectra of all observations were fitted simultaneously over the range $0.5-7~$keV with the absorbed thermal plasma emission model \textsc{phabs(apec)} \citep[][]{balucinska-church_photoelectric_1992,smith_collisional_2001}. To use this model, we need the redshift ($z = 0.234$) and the Galactic absorption along the line of sight (\citealt[][]{kalberla_leidenargentinebonn_2005}, $n_H = 3.0 \times 10^{20}~$atoms~cm$^{-2}$). Additionally, this model has three free parameters: the temperature (in keV), the metallicity and the normalisation. The metallicity of the gas is a fraction of the solar abundance determined by \citet{anders_abundances_1989} for comparison with previous X-ray studies of clusters. The best-fit spectral model is found by minimizing the C-statistic \citep[][]{cash_parameter_1979}, used because of the low count rates.

Maps are created using the best-fit values for each parameter in each region. The temperature map for a signal-to-noise of 32, and therefore regions of $\sim 1000~$counts, is presented in the upper-right panel of Fig.~\ref{fig:Counts_Tmap_Filament}. The uncertainties on the temperature in the subcluster core are on average 10\%, but increase to 12\% in the tail and  40\% in the ambient regions. This is due to the presence of the Fe-L line in spectra at lower temperatures which facilitates the differentiation of the temperature values. Temperatures in the subcluster core are around $kT \sim 2-4~$keV and increase to $kT \sim 4-6~$keV in the ram-pressure stripped tail, still colder than the average ICM temperature ($kT \sim 8-10~$keV). We can see colder plumes in the ram-pressure stripped tail, but most importantly, we can see a plume of colder gas emanating from the east of the subcluster core, clearly discernible from the hotter ambient gas. The temperatures along the plume are between $kT \sim 2.1-5.6~$keV. This plume is highlighted by a conic region in the panels of Fig.~\ref{fig:Counts_Tmap_Filament}, and is the same as the plume shown in Fig.~\ref{fig:unsharpmask_optical}. We created a conic region over the plume, split in 2D in trapezoid regions, to represent a partial cone of gas surrounding the eastern side of the core. The other plumes in the ram-pressure stripped tail are confused by projection.

We tested the robustness of the plume structure by creating multiple temperature maps with different signal-to-noise ratios and varying constraints on the sizes of the regions. The one presented in Fig.~\ref{fig:Counts_Tmap_Filament} was the one with the best middle-ground between small regions and acceptable uncertainties, however, we could clearly see the plume in all of them and the temperature gradients were consistent. We therefore conclude that this plume is a distinct structure, and study thermal conduction along it. Even if there is still contact with the main body of the subcluster core on the west side, conduction from the surrounding hot gas will dominate the heating given its temperature and the strong temperature dependence of conductivity.

\subsection{Temporal evolution}\label{sec:results_profilecuts}

\subsubsection{Parameters of the model}
\begin{figure*}
\centering
\begin{minipage}[c]{1.0\linewidth}
\centering 
\includegraphics[width=1.0\textwidth]{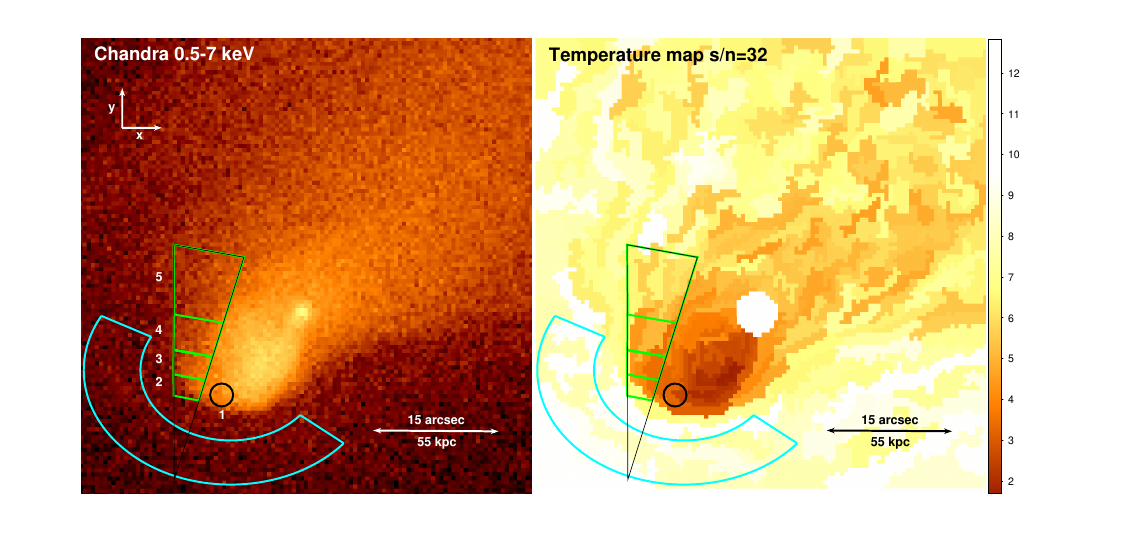}
\includegraphics[width=0.5\textwidth]{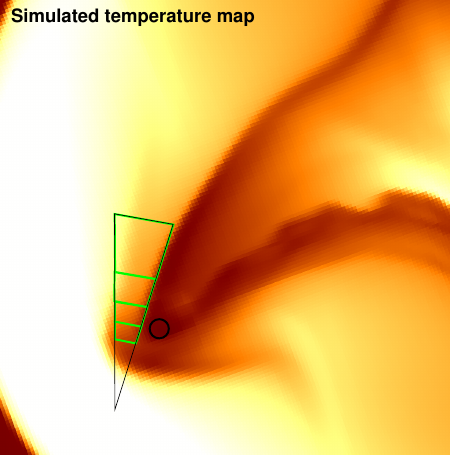}
\end{minipage}
\caption[]{\textbf{Upper-left:} Exposure-corrected X-ray image of Abell~2146 in the $0.5-7~$keV energy band (in counts~cm$^{-2}$~s$^{-1}$~pixel$^{-1}$)  \textbf{Upper-right:} Projected temperature map (keV). Each region has a signal-to-noise of 32 ($\sim1000~$counts per regions). The white circle represents the excluded point source due to the AGN. \textbf{Bottom:} Simulated temperature map created following \citet{chadayammuri_constraining_2022}, but with a flatter central gas density profile for the subcluster core.  For all panels, the green trapezoid regions represent the different regions of the plume, described by a partial cone, and the black circle region represent the stagnation point. In the upper-left panel, the regions are numbered to correspond to the rows in Tables~\ref{tab:7regions} and \ref{tab:velocities}. The cyan region represents the background region used for the ambient \textsc{apec} component when fitting for the temperature, metallicity and norm for each region. For all panels, the north direction is up and the east direction is towards the left, while the $y$-axis is in the northern direction and the $x$-axis in the western direction.}
\label{fig:Counts_Tmap_Filament}
\end{figure*}

In this section, we will look at how quickly the temperature in the plume would rise if the heating was only due to conduction. To study the change in temperature along the plume, and therefore the effect of  conduction from the ambient hot gas that heats the cool gas of the plume, we separated it into different regions. In Fig.~\ref{fig:Counts_Tmap_Filament}, 4~different regions are overlaid in green on the X-ray image (on the upper-left panel), on the temperature map with regions of signal-to-noise of $\textrm{S/N} \sim 32$ (on the upper-right panel) and on the simulated temperature map (on the bottom panel), and trace the partial cone of gas. The circular black region represents the position of a stagnation point of the flow. The cyan region represents the ambient, hot gas and is upstream of the cool core break up, therefore should not have been affected by any of the subcluster core gas. The gas temperature at this position is $kT_{\rm hot}=9.6 \substack{+0.6 \\ -0.5}$~keV, determined by using \textsc{xspec} to jointly fit an absorbed \textsc{apec} plasma model to each of the 75 separate $0.5-7$~keV spectra of the observations, leaving the temperature, abundance and normalisation free to vary. 

The break up of the subcluster core produces irregular clumps of cool gas that cannot be approximated as symmetrical, precluding the  use of deprojection to determine the properties of the plume's gas \citep[][]{fabian_x-ray_1980}.  Instead, we have used a model with two thermal components, one to account for the properties of the cool plume (plume \textsc{apec} component) and the other for the hot ambient gas surrounding it in 3D (ambient \textsc{apec} component). For each region on the plume, we fitted an absorbed double \textsc{apec} plasma model to the separate $0.5-7$~keV spectra of each of the 75 observations.  For the ambient \textsc{apec} component, we fixed the temperature ($kT_{\rm hot}=9.6 \substack{+0.6 \\ -0.5}$~keV), abundance ($Z_{\rm hot}=0.24 \pm 0.07$) and normalisation (${\rm norm}_{\rm hot}=(6.7 \pm 0.11) \times 10^{-5}$~\textsc{xspec}~units, for the background region) using the values found previously for the hotter gas in the cyan region.  The latter parameter was scaled by the difference in sizes of the plume regions compared to the size of the background region.

The normalisation and temperature for each plume \textsc{apec} component were left free to vary and the abundance was set at 0.5~Z$_{\odot}$. We fixed the abundance as it was found to be consistent for all regions. For all regions, we used a redshift of $z=0.234$ and an equivalent hydrogen column density of $n_H=3.0 \times 10^{20}$~cm$^{-2}$.

For each region, we also looked at the electron density, pressure, mass and mass/kpc$^2$ of the gas. With the normalisation found in \textsc{xspec}, and considering spherical (circle in 2D) or conical (trapezoid in 2D) regions accordingly, we found the electron density and therefore the gas density. From the gas density, the gas mass of each region could be found and therefore the mass/kpc$^2$, as the masses cannot simply be compared as the regions used become larger along the plume. A proxy for the pressure of each region was also found by multiplying their electron density by their temperature to see the trend of it. In Table~\ref{tab:7regions}, the electron gas density of each region, their mass per kpc$^2$ and the proxy for their pressure is shown. 

 \begin{table*}
\caption{\textbf{Conduction from the hot ambient gas.} Values needed to study the conduction fluxes from the hot ambient gas in the plume split into 5 regions, following Fig.~\ref{fig:Counts_Tmap_Filament}. The columns are: 1. Region number identified in the upper-left panel of Fig.~\ref{fig:Counts_Tmap_Filament}, region 1 being the most westerly region and stagnation point; 2. Speed of the gas in the $x-y$ plane relative to the stagnation point, in km~s$^{-1}$; 3. Temperature of the gas ($kT$) from the fit in \textsc{xspec}, in keV; 4. Electron gas density of each region, in electrons~cm$^{-3}$; 5. Proxy for the pressure of the gas in each region ($P \propto n_e \times kT$), in keV~electrons~cm$^{-3}$; 6. Mass of the gas in each region per kpc$^{2}$, in $10^7~$M$_{\odot}~$kpc$^{-2}$; 7. Required conduction flux for a region, following Eq.~\ref{Eq:q-}, in $10^{-3}$~erg~s$^{-1}$~cm$^{-2}$; 8. Available conduction flux from the ambient hot gas, following Eq.~\ref{Eq:q+}, in $10^{-3}$~erg~s$^{-1}$~cm$^{-2}$; 9. Saturation flux, following Eq.~\ref{Eq:qsat}, in $10^{-3}$~erg~s$^{-1}$~cm$^{-2}$; 10. Ratio between the available flux and the required one.}
  \label{tab:7regions}
  \resizebox{17.7cm}{!}{%
  \begin{tabular}{cccccccccc}
    \hline
    \thead{Regions} & \thead{$v$} & \thead{$kT$} & \thead{$n_e$} & \thead{$kT \times n_e$} & \thead{$m/$kpc$^{2}$} & \thead{$q^{-}$} &\thead{$q^{+}$} & \thead{$q_{sat}$} & \thead{Ratio}  \\
    \hline

1 & $0$ & $2.05 \substack{+0.10 \\ -0.09}$ & $0.0350 \substack{+0.0006 \\ -0.0005}$ & $0.072 \substack{+0.004 \\ -0.003}$ & $1.216 \pm 0.019$ & NA & $348 \substack{+5 \\ -4}$ & $280 \substack{+30 \\ -20}$ & NA \\ 
2 & $229$ & $2.4 \substack{+0.3 \\ -0.2}$ & $0.0488 \substack{+0.0013 \\ -0.0014}$ & $0.117 \substack{+0.014 \\ -0.011}$ & $1.22 \pm 0.03$ & $0.6 \substack{+0.5 \\ -0.4}$ & $267 \substack{+10 \\ -8}$ & $660 \pm 60$ & $459 \substack{+385 \\ -304}$ \\ 
3 & $231$ & $3.1 \pm 0.2$ & $0.0520 \substack{+0.0010 \\ -0.0009}$ & $0.161 \substack{+0.012 \\ -0.011}$ & $1.61 \pm 0.03$ & $1.4 \substack{+0.7 \\ -0.6}$ & $195 \substack{+7 \\ -6}$ & $600 \substack{+60 \\ -50}$ & $143 \substack{+73 \\ -62}$ \\ 
4 & $272$ & $3.35 \substack{+0.22 \\ -0.19}$ & $0.0410 \pm 0.0006$ & $0.137 \substack{+0.009 \\ -0.008}$ & $1.60 \pm 0.02$ & $0.4 \substack{+0.5 \\ -0.4}$ & $149 \pm 5$ & $530 \substack{+50 \\ -40}$ & $396 \substack{+507 \\ -464}$ \\ 
5 & $430$ & $5.0 \pm 0.3$ & $0.0286 \pm 0.0003$ & $0.144 \substack{+0.009 \\ -0.008}$ & $1.512 \substack{+0.017 \\ -0.016}$ & $2.0 \pm 0.4$ & $81 \pm 5$ & $460 \pm 40$ & $41 \substack{+10 \\ -9}$ \\

        \hline
  \end{tabular}%
  }
 \end{table*}

To estimate the required conductive heat flux from the hot ambient gas into the cool plume and the available conductive heat flux from the ICM, we approximate the plume as a partial cone divided into trapezoidal segments each moving at a speed $v$ along the plume.

\subsubsection{Velocities from the simulation}
The velocities used are from the simulations of the gas flow in A2146 (see \citealt{chadayammuri_constraining_2022}). This study used a suite of idealized hydrodynamical simulations of binary cluster mergers to determine the merger configuration, cluster masses, velocities, impact parameter and orientation of A2146.  These simulations constrained the merger scenario and successfully reproduced the large-scale ICM structure.  In particular, the gas velocity behind the bow shock from \citet{russell_structure_2022} is a key measurement from the X-ray observations that constrains the velocities around the subcluster core, as it determines the incoming speed of the gas. In the simulations, there is a non-zero impact parameter, meaning that the system has net angular momentum. Additionally, the simulation included self-gravity of the gas and the subcluster core was initialised with a cool core density and temperature profile. The super-NFW profile \citep{lokas_properties_2001} was used to model the initial dark matter distribution of a cluster and the gas was set up in hydrostatic equilibrium with the dark matter, as well as modeling the gas density with the modified beta profile of \citet[][]{vikhlinin_chandra_2006}. However, it does not include dark matter or gas to represent a BCG in either subcluster. Indeed, the goal was to reproduce the large-scale X-ray structure, thus finer details in the core were not captured. Therefore, for the purpose of our study, the simulation was modified with a shallower central gas density profile for the subcluster core, which ensured that the gas was less bound, thus more prone to disruption, and reproduced the stripped structure.
It is important to note that the simulations of \citet{chadayammuri_constraining_2022} are hydrodynamic simulations, i.e. they do not include magnetic fields or thermal conduction. As a result, they are not expected to perfectly reproduce the observations, especially in regions like cold fronts where magnetic draping and thermal conduction are expected to be important. Nevertheless, the subcluster core with the ram-pressure stripped gas behind it is resolved and we see plumes forming on both sides of the subcluster core, as in the observations. 

The inclusion of a magnetic draping layer would not necessarily inhibit gas stripping \citep[e.g.][]{ruszkowski_impact_2014} but it may alter the gas velocities along the plume.  Our results are robust to modest variations in the gas velocities and we place strong constraints on conduction even up to the maximum postshock gas velocity of $1100 \pm 100~$km~s$^{-1}$.  The velocities of the gas in the eastern plume of the simulated subcluster provide a clear improvement over the maximum velocity for our analysis. The temperature map of this version of the simulations is shown in the bottom panel of Fig.~\ref{fig:Counts_Tmap_Filament}.

The trajectory of the core is slightly different in the observations and \citet{chadayammuri_constraining_2022} simulations, hence we used the stagnation point at the leading edge of the subcluster core in the observations and the simulations as a point of reference, as well as the flow direction of the ram-pressure stripped core to align the simulations. This is what is represented in Fig.~\ref{fig:Counts_Tmap_Filament}. The stagnation point was also used to place the regions on the simulations.

The values of the velocity $v$ are relative to the velocity of the stagnation point at the leading edge of the simulated subcluster core ($v_{\rm stagnation (x,y)} = (1319,395)~$km~s$^{-1}$). This way, the frame of reference is taken into consideration when simulating the relative velocity of the plume (see Table~\ref{tab:velocities} for the relative velocity values of each region in each direction, the $x$-axis pointing to the west and $y$-axis to the north).  Note that $v_{\rm z}$ is negligible in this case because the merger axis is close to the plane of the sky and thus we are only using the $x-y$ plane for the relative velocity vector. Indeed, the hydrodynamical simulations of A2146 show that there is only minimal differences between temperature profiles for inclination angles of $<20\degr$ \citep[][]{chadayammuri_constraining_2022}. The velocities in Table~\ref{tab:velocities} are simply the mean of the gas velocities in each region, sampled along a uniform grid of resolution 3.4~kpc. The uncertainties in the $x$-axis from that spread are between $1.8\%$ and $8.5\%$.
 
 \begin{table}
  \caption{Relative velocities of each region from hydrodynamic simulations similar to the ones from \citet{chadayammuri_constraining_2022} of the gas flow in A2146. The columns are: 1. Region number identified in the upper-left panel of Fig.~\ref{fig:Counts_Tmap_Filament}, region 1 being the most westerly region and stagnation point; 2. \& 3. Relative velocity in the $x$ and $y$ directions of the gas, relative to the stagnation point. The $x$-axis points to the west and $y$-axis to the north; 4. Relative velocity of each regions in the $x-y$ plane, relative to the velocity of the stagnation point at the leading edge of the simulated subcluster core.}
  \label{tab:velocities}
  \begin{tabular}{cccc}
    \hline
    \thead{Regions} & \thead{$v_x$} & \thead{$v_y$} & \thead{$v$} \\
    \thead{} & \thead{[km~s$^{-1}$]} & \thead{[km~s$^{-1}$]} & \thead{[km~s$^{-1}$]} \\
    \hline

1 & $0$ & $0$ & $0$ \\ 
2 & $154$ & $-170$ & $229$ \\ 
3 & $-73.0$ & $219$ & $231$ \\ 
4 & $49.8$ & $268$ & $272$ \\ 
5 & $145$ & $405$ & $430$ \\ 

        \hline
  \end{tabular}
 \end{table}

\subsubsection{Constraints on conduction}

To derive the limits on the role of conduction, we treat the plume as a cone, represented in two-dimension by the trapezoids shown in Fig.~\ref{fig:Counts_Tmap_Filament} and suppose that the temperature increase of the plume is entirely due to conduction from the hotter surrounding plasma. We assume that the gas of the core is flowing along the plume as if it was flowing through a pipe. If we focus on a parcel of gas of length $l_i$ and radius $r_i$ going through the flow, the relevant energy equation (ignoring radiative losses) is given by

\begin{equation}
   \frac{1}{K} \frac{DK}{Dt} = - \frac{(\gamma - 1)}{P} \nabla \cdot q^-,
\end{equation}
where $K=P/\rho^{\gamma}$ is the entropy index with $\ln (K)$ being proportional to the entropy of the gas, $\rho$ is the gas density, $\frac{D}{Dt}$ is the usual convective derivative, $P$ is the gas pressure, $q^-$ is the conductive heat flux and $\gamma$ is the ratio of the specific heat capacities (taken to be $\gamma=5/3$ for this fully ionized plasma). As $P \propto \rho T$ for ideal gas, where $T$ is the temperature of the gas, $1/\rho^{\gamma} \propto (T/P)^{\gamma}$ and therefore $K=P/\rho^{\gamma} \propto P^{1-\gamma}T^{\gamma}$. Thus,

\begin{equation}
    \frac{1}{K} \frac{DK}{Dt} = \frac{D}{Dt} (\ln K) = \frac{(1-\gamma)}{P} \frac{DP}{Dt} + \frac{\gamma}{T} \frac{DT}{Dt}.
\end{equation}

In Table~\ref{tab:7regions}, we see that the pressure remains fairly constant along the plume, and therefore we can approximate that $\frac{DP}{Dt} = 0$. The temperature indeed doubles from region 2 to region 5, while there is only an increase of $20\%$ of the pressure between those regions. Hence, ignoring pressure changes, we have

\begin{equation}
    \frac{1}{K} \frac{DK}{Dt} = \frac{\gamma}{T} \frac{DT}{Dt} = - \frac{(\gamma - 1)}{P} \nabla \cdot q^-.
\end{equation}

As $P=n_{\rm tot}kT$, with $kT$ the temperature and $n_{\rm tot}$ the total particle number density,

\begin{equation}
    \gamma n_{\rm tot}k \frac{DT}{Dt} = - (\gamma - 1) \nabla \cdot q^-.
\end{equation}

As we are looking at the change in temperature that happens between a small increment of time $\Delta t$, we can approximate this equation to

\begin{equation}
    \gamma n_{\rm tot}k \frac{\Delta T}{\Delta t} = - (\gamma - 1) \nabla \cdot q^-,
\end{equation}
where we have replaced the derivative with the ratio of finite changes. We can integrate over the volume ($V$) of a parcel, which gives us

\begin{equation}
    \frac{\gamma}{\gamma -1} Vn_{\rm tot}k \frac{\Delta T}{\Delta t} = A_{\rm surface} q^-,
\end{equation}
where $A_{\rm surface}$ is the surface area of the plume. Based on the chemical composition of the ICM (hydrogen, helium and a small fraction of metals) and the solar abundance of each element, we approximate $n_{\rm tot} = 1.92n_e$, the electron number density. For each segment $i$ of the cone, $n_{e, i}$ is its electron density (in cm$^{-3}$), $kT_i$ its temperature (in keV) and $V_i$ and $A_{{\rm surface}, i}$ are its volume (in cm$^3$) and surface area (in cm$^2$) respectively. The time it takes for gas to travel from one end of a segment to the other is $\Delta t=l_i/v_i$, where $v_i$ is the velocity of the gas in that segment.  Therefore, the required conduction heat flux at the boundary of each segment $i$ of the cone is

\begin{equation}\label{Eq:q-}
\begin{split}
  q\substack{- \\ i} & = \frac{5}{4} \frac{V_i}{A_{{\rm surface}, i}} 1.92n_{e, i} k\Delta T \frac{v_i}{l_i} \\
  & = 2.4 n_{e,i} k(T_i-T_{i-1}) \frac{v_i}{l_i} \frac{V_i}{A_{{\rm surface}, i}}.
\end{split}
\end{equation}

Our plume is described by a cone split in 2D in trapezoids. For each region, $l_i$ in Fig.~\ref{fig:Counts_Tmap_Filament} is the height of the trapezoid, $2a_i$ is the length of its shorter base and $2b_i$ of its longer base. From this, $r_i=\frac{a_i + b_i}{2}$ is the distance between the centre of the trapezoid and the ambient gas. For Eq.~\ref{Eq:q-}, we need to find the volume and surface area of partial cones. Subtracting a small cone of radius $a_i$ from a bigger cone of radius $b_i$, we get for the partial cones (representing the trapezoids in 3D) 

\begin{equation}
    V_i = \frac{1}{3} \pi l_i (b_i^2 +a_ib_i + a_i^2),
\end{equation}

\begin{equation}
    A_{{\rm surface}, i} = \pi (a_i + b_i) \sqrt{l_i^2 + (b_i-a_i)^2}.
\end{equation}

The plume is likely a partial cone of gas just outside the eastern side of the cool core or a little surrounding it, therefore we use only 7/8$^{\rm th}$ of the surface area of the cone in our calculations. Although the available conduction flux is dependent on the chosen geometry, we tested numerous potential geometries and only found variations of order unity, therefore producing the same conclusions.  We proceed with the partial cone geometry as it provides a simple yet accurate description of the observational data. 

The available conduction flux into the side of the cone of gas is given by \citep[][]{spitzer_physics_1956,cowie_evaporation_1977,sarazin_x-ray_1988,ettori_chandra_2000}
\begin{equation}\label{Eq:q+}
q\substack{+ \\ i} = \kappa \frac{k(T_h-T_i)}{r},
\end{equation}
where $kT_h$ is the temperature of the hot ambient gas in the background region and $\kappa$ is given by
\begin{equation}
    \kappa = 8.2 \times 10^{20} \left(\frac{kT_h}{10~\textrm{keV}}\right)^{5/2} \textrm{erg~s$^{-1}$~cm$^{-1}$~keV$^{-1}$}.
\end{equation}

The uncertainty on $\kappa$ is not taken into account while calculating the uncertainty for each region on the available conduction flux, because this value is a constant value between regions, applied to all of them.

The heat flux will saturate to the limiting value that can be carried by the electrons \citep[][]{cowie_evaporation_1977}, and therefore this value needs to be computed and compared to the available conduction flux. This saturation flux is given by
\begin{equation}\label{Eq:qsat}
    q_{sat} = 0.023 \left(\frac{kT_h}{10~\textrm{keV}}\right)^{3/2} \left(\frac{n_{e,h}}{10^{-3}~\textrm{cm$^{-3}$}}\right) \quad \textrm{erg~s$^{-1}$~cm$^{-2}$},
\end{equation}
where $n_h$ is the density of the hot ambient gas (in cm$^{-3}$). The values for those three conduction fluxes for each region of the plume can be found in Table~\ref{tab:7regions}. This table also gives the relative velocity, temperature, electron density, proxy for the pressure and mass per kpc$^2$ of the gas of each region, and the ratio between the available flux and the required one, allowing us to study the possible suppression of conduction.

Looking at Table~\ref{tab:7regions}, we can see that the saturation flux is larger than the available conduction flux in the ambient hot gas for every region, and therefore does not constrain the available flux. The heat flux available due to conduction at the Spitzer rate would also be orders of magnitude larger than the heat flux required to account for the temperature rise along the plume. The slow temperature rise therefore requires the thermal conductivity to be suppressed by at least two orders of magnitude below the Spitzer value, as can be seen by the ratio between the available and required conduction heat flux. We also clearly see that the temperatures increase along the plume, as well as their uncertainties. Finally, the gas pressure of each region in the plume is roughly constant along the plume, which follows our assumption for the conduction equations used in this section. However, our analysis has an implicit assumption of steady flow, as for our calculations we assume that the properties of the fluid element entering each region are the same as those of the element exiting the preceding region. The results are unlikely to differ by more than a factor of a few, thus the conclusions are not affected. In the next section, we will go into more detail on the significance of these results. 

\section{Discussion}\label{sec:discussion}

\subsection{Suppression of conductivity}

Following the earlier studies of cold fronts, filaments and small scale structures in the temperature map \citep[e.g.][]{ettori_chandra_2000,fabian_chandra_2001,markevitch_chandra_2003}, we found that the heat flux, and therefore the thermal conduction, must be suppressed by a factor of at least two orders of magnitude for the plume in A2146 to exist, as the energy used is about 1\% of the energy available. Earlier studies were limited by the approximations used to determine the survival time of the structure (i.e. dynamical or crossing time). In our study, the plume is spatially resolved, and we use the result of a detailed computer simulation  tailored for A2146 (see \citealt[][]{chadayammuri_constraining_2022}) to determine the velocities of the gas, which with the size of each adjacent segment of the plume, yields the time scale. 

Suppression could be explained by magnetic draping around the structures in the ICM \citep[e.g.][]{lyutikov_magnetic_2006} due to highly tangled magnetic fields \citep[e.g.][]{chandran_thermal_1998,carilli_cluster_2002,govoni_magnetic_2004} or to strong untangled magnetic fields perpendicular to the change in temperature and density \citep[e.g][]{vikhlinin_chandra_2001,asai_three-dimensional_2005,dursi_draping_2008}, as transport processes, including conduction, are reduced perpendicular to the direction of the magnetic field. When a cold front moves in a magnetized medium, the flow will stretch and drape the field lines around the interface and the field will amplify and align with the cold front \citep[][]{zuhone_cold_2016}. \citet{chadayammuri_turbulent_2022} studied the consequences of magnetic fields of different strengths on the X-ray features seen in observations through magnetohydrodynamic simulations of A2146 using a tangled initial magnetic field. They found that the tangled magnetic field gets the most amplified around the cold front at the edge of the subcluster core. 

Supporting the impact of strong untangled magnetic fields on the suppression of conduction, \citet{meinecke_strong_2022} made a laboratory replica of the turbulence in the plasma of galaxy clusters and found a reduction of heat conduction of at least two orders of magnitude from the Spitzer's values. In this study, they used a laser laboratory experiment to replicate a high-$\beta$ turbulent plasma with weakly collisional and magnetised electrons resembling the conditions of galaxy clusters, where $\beta$ is the ratio of the electron thermal pressure to the magnetic pressure and has a value of 44. It was the first direct experimental evidence of such a suppression in those conditions.

Interestingly, the conduction suppression found here is also similar to the suppression due to the whistler scattering found by \citet{roberg-clark_suppression_2018}. This mechanism suppresses heat flux parallel to magnetic fields because high heat fluxes generate whistler waves that scatter the electrons, suppressing their ability to transport heat. The thermal heat transport quickly saturates at a level of $\sim q_{\rm sat}/\beta$, where the ratio of gas to magnetic pressure is $\beta\sim 100$ \citep[][]{drake_whistler-regulated_2021}.

We cannot be sure what is the origin of the temperature increase along the plume, as there could be many phenomena explaining this heating. The temperature of the gas within the plume may also change due to mixing \citep[e.g.][]{nulsen_transport_1982,begelman_turbulent_1990,tonnesen_its_2021}. This would vastly increase the surface area of the contact region between hot and cold gas and so require much larger suppression of conduction than estimated here. The unknown magnetic field configuration prevents a more detailed consideration of mixing. However, conduction at the Spitzer rate would heat up the gas directly. Suppression of the conduction therefore needs to take place, whatever occurs regarding other heating mechanisms. Furthermore, they would increase the temperature of the plume, implying even more suppression of the conduction needed. Thus, we find here a conservative limit for it.

\subsection{Conduction along the plume}
        
In the past, a number of simulations have allowed for anisotropic thermal conduction, using the Spitzer  conductivity parallel to the magnetic field lines \citep[e.g.][]{zuhone_cold_2013}. However, suppression could still exist along the field lines and therefore the plume in our case (e.g. via whistler mode, \citealt{roberg-clark_suppression_2018}). We do not know the orientation of the magnetic field lines in A2146, but based on the high suppression of the conduction between the plume and the hot ambient gas, the magnetic field should run along the interface of the plume and the hot gas. 

We have checked whether the temperature gradient could itself produce the conduction necessary. We looked at a parcel of gas moving along the plume, but with heat coming down from heat pipes connecting the hotter plasma of region 5 to the colder plasma of region 2 next to the cool core. In Table~\ref{tab:conduction}, for each region, we put the temperature, required conduction flux ($q^{-}$) and available one from the hot ambient gas ($q^{+}$) from Table~\ref{tab:7regions}. We also added a column with the available conduction flux between segments of the plume looking at the conduction between two adjacent regions ($q^{+}_a$). We then calculated the fraction of the conduction along the plume versus the one from the hot gas ($\frac{q^{+}_a}{q^{+}}$, in percentage) and the ratio between the available flux along the plume and the required one (Ratio$_a$). We also decided to look at those three values again, but for the conduction only between the first and last regions of the partial cone ($q^{+}_{a,ext}$, $\frac{q^{+}_a}{q^{+}}_{ext}$, Ratio$_{a,ext}$). It seems that the conduction along the plume needs to be suppressed, but the uncertainties are too great, and we thus have no strong evidence for suppression along the plume. However, \citet{roberg-clark_suppression_2016} show that for flow along the magnetic field, the conductive heat flux saturates at a much lower value than estimated by \citet{cowie_evaporation_1977} due to excitation of whistlers. This could explain the low conduction along the field.  In addition, we found that the values of conduction available along the plume are between 0.15\% and 5.43\% of the available conduction flux from the hot ambient gas. For those reasons, it can be safely ignored.

\begin{table*}
\caption{\textbf{Conduction along the plume.} Values needed to study the conduction fluxes along the plume split into 5 regions, following Fig.~\ref{fig:Counts_Tmap_Filament}. The columns are: 1. Region number identified in the upper-left panel of Fig.~\ref{fig:Counts_Tmap_Filament}, region 1 being the most westerly region and stagnation point; 2. Temperature of the gas ($kT$) from the fit in \textsc{xspec}, in keV; 3. Required conduction flux for a region, following Eq.~\ref{Eq:q-}, in $10^{-3}$~erg~s$^{-1}$~cm$^{-2}$; 4. Available conduction flux from the ambient hot gas, following Eq.~\ref{Eq:q+}, in $10^{-3}$~erg~s$^{-1}$~cm$^{-2}$; 5. Available conduction flux from the adjacent, hotter segment of the plume, following Eq.~\ref{Eq:q+}, in $10^{-3}$~erg~s$^{-1}$~cm$^{-2}$; 6. Fraction of the conduction along the plume versus the one from the hot gas, in percentage; 7. Ratio between the available flux along the plume and the required one; 8., 9. \& 10. Same than column 5., 6. \& 7. respectively, but for the conduction fluxes from region 5 into region 2.}
\label{tab:conduction}
\resizebox{17.7cm}{!}{%
\begin{tabular}{cccccccccc}
\hline
\thead{Regions} & \thead{$kT$} & \thead{$q^{-}$} & \thead{$q^{+}$} & \thead{$q^{+}_a$} & \thead{$\frac{q^{+}_a}{q^{+}}$} & \thead{Ratio$_a$} &\thead{$q^{+}_{a,ext}$} & \thead{$\frac{q^{+}_a}{q^{+}}_{ext}$} & \thead{Ratio$_{a,ext}$}  \\
\hline

1 & $2.05 \substack{+0.10 \\ -0.09}$ & NA  & $348 \substack{+5 \\ -4}$& $0.5 \substack{+0.4 \\ -0.3}$  & 0.15  & NA  &  &  &  \\ 
2 & $2.4 \substack{+0.3 \\ -0.2}$ & $0.6 \substack{+0.5 \\ -0.4}$ & $267 \substack{+10 \\ -8}$& $1.6 \substack{+0.8 \\ -0.7}$  & 0.58  & $2.68 \substack{+2.63 \\ -2.11}$  & $5.2 \substack{+0.8 \\ -0.7}$ & 7.34 & $8.9 \substack{+7.6 \\ -6.0}$ \\ 
3 & $3.1 \pm 0.2$ & $1.4 \substack{+0.7 \\ -0.6}$ & $195 \substack{+7 \\ -6}$ & $0.5 \substack{+0.7 \\ -0.6}$ & 0.27 & $0.39 \substack{+0.54 \\ -0.49}$  &   &  &  \\ 
4 & $3.35 \substack{+0.22 \\ -0.19}$ & $0.4 \substack{+0.5 \\ -0.4}$ & $149 \pm 5$ & $8.1 \substack{+1.8 \\ -1.6}$  & 5.42  & $21.5 \substack{+27.9 \\ -25.5}$  &   &  &  \\ 
5 & $5.0 \pm 0.3$ & $2.0 \pm 0.4$ & $81 \pm 5$ & NA  & NA  & NA  &   &  &  \\ 

       \hline
  \end{tabular}%
  }
 \end{table*}

\section{Summary}\label{sec:conclusion}


The new 2~Ms $\textit{Chandra}$ observations of the cluster merger A2146 allowed us for the first time to study conduction processes in the ICM of galaxy clusters using a tailor made model to determine the flow velocities. A2146 is the merger of two cool core clusters. The subcluster core was ram-pressure stripped during the merger, which created a tail of cool gas behind the core (see Fig.~\ref{fig:A2146image}). The resolution of the new data allowed us to see a clear cool plume coming from the core distinguishable from the hot ambient gas, thought to be a partial cone with the section closest to the core missing (see Fig.~\ref{fig:unsharpmask_optical}). From the detailed temperature map of the plume and the gas velocities along a comparable structure in the hydrodynamical simulations (similar to the ones in \citealt{chadayammuri_constraining_2022}), we studied the survival of the plume and suppression of conduction in the cool core.

By mapping the temperature gradient along the plume (see Fig.~\ref{fig:Counts_Tmap_Filament}), we calculated the required conduction flux to explain the change in temperature over a series of sections along the plume's length. Those values were then compared to the available conduction flux in the hot ambient gas following the Spitzer rates. By comparing those two fluxes, we found that there is a suppression of the conduction compared to the Spitzer rates of at least two orders of magnitude depending on the region, our results being lower limits. This could be explained by magnetic draping due to highly tangled magnetic fields or strong untangled magnetic fields along the plume. Therefore, our results show, for the first time using measurements of the geometry and time scale of the plume rather than using arguments of the survival time of the structures, that conduction is indeed strongly suppressed in some regions of galaxy clusters. 

\section*{Acknowledgements}

We would like to thank the anonymous referee for providing constructive comments and raising interesting points that helped in improving the quality and presentation of the manuscript. ARL is supported by the Gates Cambridge Scholarship, by the St John's College Benefactors' Scholarships, by NSERC through the Postgraduate Scholarship-Doctoral Program (PGS D) under grant PGSD3-535124-2019 and by FRQNT through the FRQNT Graduate Studies Research Scholarship - Doctoral level under grant \#274532. HRR acknowledges support from an STFC Ernest Rutherford Fellowship and an Anne McLaren Fellowship.  ACE acknowledges support from STFC grant ST/P00541/1.

\section*{Data Availability}

The data underlying this article are available in the \textit{Chandra} and \textit{Hubble Space Telescope} archives and in this article. The observation identification numbers for the \textit{Chandra} observations of A2146 are in \citet{russell_structure_2022}. The \textit{Hubble Space Telescope} observation in Fig.~\ref{fig:A2146image} is the visit \#01 of proposal \#12871 with the F606W filter. The reduced images, temperature maps and simulations in Figs.~\ref{fig:A2146image}, \ref{fig:unsharpmask_optical} and  \ref{fig:Counts_Tmap_Filament} generated for this research could be shared on reasonable request to the corresponding author.




\bibliographystyle{mnras}
\bibliography{ref_PhD_A2146_core} 




\appendix


\bsp	
\label{lastpage}
\end{document}